\documentclass[letter,11pt]{article}
\usepackage[top=1in,bottom=1in,left=1in,right=1in]{geometry}
\usepackage{comment}

\usepackage{slashed}
\usepackage{epsfig}
\usepackage{comment}
\usepackage{cancel}
\usepackage{bbm}
\usepackage{array}
\usepackage{bigints}
\usepackage{booktabs}
\usepackage{color}
\usepackage{dsfont}
\usepackage{float}
\usepackage{framed}
\usepackage{graphicx}
\usepackage{indentfirst}
\usepackage{mathrsfs}
\usepackage{multirow}
\usepackage{simpler-wick}
\usepackage{subdepth}
\usepackage{titlesec}
\usepackage[dotinlabels]{titletoc}
\usepackage{wrapfig}
\usepackage[all]{xy}
\usepackage[vcentermath]{youngtab}
\usepackage{relsize}
\usepackage{hyperref}
\usepackage{xcolor}
\usepackage{soul}
\numberwithin{equation}{section}

\usepackage[utf8]{inputenc}
\usepackage{slashed}
\usepackage{amsmath}
\usepackage{amsfonts}
\usepackage{amssymb}
\usepackage{cite}
\usepackage{mathtools}

\usepackage{cleveref}
\crefname{section}{§}{§§}
\crefname{section}{§}{§§}

   \makeatletter
  \let\over=\@@over \let\overwithdelims=\@@overwithdelims
  \let\atop=\@@atop \let\atopwithdelims=\@@atopwithdelims
  \let\above=\@@above \let\abovewithdelims=\@@abovewithdelims
\renewcommand\section{\@startsection {section}{1}{\z@}%
{-3.5ex \@plus -1ex \@minus -.2ex}
{2.3ex \@plus.2ex}%
{\normalfont\large\bfseries}}

\renewcommand\subsection{\@startsection{subsection}{2}{\z@}%
{-3.25ex\@plus -1ex \@minus -.2ex}%
{1.5ex \@plus .2ex}%
{\normalfont\bfseries}}

\makeatletter
\DeclareFontFamily{OMX}{MnSymbolE}{}
\DeclareSymbolFont{MnLargeSymbols}{OMX}{MnSymbolE}{m}{n}
\SetSymbolFont{MnLargeSymbols}{bold}{OMX}{MnSymbolE}{b}{n}
\DeclareFontShape{OMX}{MnSymbolE}{m}{n}{
    <-6>  MnSymbolE5
   <6-7>  MnSymbolE6
   <7-8>  MnSymbolE7
   <8-9>  MnSymbolE8
   <9-10> MnSymbolE9
  <10-12> MnSymbolE10
  <12->   MnSymbolE12
}{}
\DeclareFontShape{OMX}{MnSymbolE}{b}{n}{
    <-6>  MnSymbolE-Bold5
   <6-7>  MnSymbolE-Bold6
   <7-8>  MnSymbolE-Bold7
   <8-9>  MnSymbolE-Bold8
   <9-10> MnSymbolE-Bold9
  <10-12> MnSymbolE-Bold10
  <12->   MnSymbolE-Bold12
}{}

\let\llangle\@undefined
\let\rrangle\@undefined
\DeclareMathDelimiter{\llangle}{\mathopen}%
                     {MnLargeSymbols}{'164}{MnLargeSymbols}{'164}
\DeclareMathDelimiter{\rrangle}{\mathclose}%
                     {MnLargeSymbols}{'171}{MnLargeSymbols}{'171}
\makeatother

\begin{document}
\begin{titlepage}
\unitlength = 1mm
\ \\
\vskip 3cm
\begin{center}

{\LARGE{\textsc{Light-ray Operators and the ${\rm w}_{1+\infty}$ Algebra}}}

\vspace{1.25cm}
Elizabeth Himwich$^{*\star}$ and Monica Pate$^{ \dagger}$

\vspace{.5cm}

$^*${\it  Princeton Center for Theoretical Science, Princeton University, Princeton, NJ 08544}\\ 
$^\star${\it  Princeton Gravity Initiative, Princeton University, Princeton, NJ 08544}\\ 
$^\dagger${\it  The Center for Cosmology and Particle Physics, New York University, New York, NY 10003}\\ 

\vspace{0.8cm}

\begin{abstract}

A universal class of light-ray operators formed from null integrals of the stress tensor is constructed in generic interacting Lorentzian conformal field theories in four spacetime dimensions. This class of light-ray operators generates the wedge algebra of ${\rm w}_{1+\infty}$, which was recently identified among the asymptotic symmetries of asymptotically flat spacetimes. In four-dimensional conformal field theories with an additional spin-one conserved current, a second universal class of light-ray operators is constructed and shown to generate the ``$S$ algebra,''  the gauge-theoretic analog of ${\rm w}_{1+\infty}$. Finally, a precise relation is established between the one-point functions of these light-ray operators in scalar states and the universal soft factors in the infinite tower of soft graviton theorems.  The results presented in this paper will be accompanied by detailed calculations and proofs in a longer forthcoming work.

\end{abstract}

\vspace{1.0cm}
\end{center}

\end{titlepage}

\pagestyle{empty}
\pagestyle{plain}

\def\vx{{\vec x}}
\def\p{\partial}
\def\po{$\cal P_O$}
\def\i{{\rm initial}}
\def\f{{\rm final}}

\pagenumbering{arabic}
 

\tableofcontents

\section{Introduction}

All relativistic quantum field theories enjoy the symmetry algebra of Poincar\'e, by definition.  However, in the absence of further specifications,  no larger symmetry algebras are generally expected to be universally shared by all relativistic quantum field theories.  Of course, additional restrictions on particle content or interactions famously lead to large classes of quantum field theories with enhanced symmetries, including conformal field theories and supersymmetric quantum field theories.  More recently, quantum field theories with propagating gauge bosons or gravitons have been shown to admit universal infinite-dimensional symmetry enhancements, which can be identified with symmetries known as ``asymptotic symmetries'' from general relativity \cite{Strominger:2017zoo}.  In gravitational theories, these symmetries organize into the algebra of ${\rm w}_{1+\infty}$. In gauge theory, the corresponding symmetry has come to be known as the ``$S$ algebra" \cite{Guevara:2021abz, Strominger:2021lvk}.

Inspired by these recent developments, in this work, we identify an avatar of these infinite-dimensional asymptotic symmetry enhancements that persists in theories \emph{without} propagating gauge bosons or gravitons.  More precisely, we identify a universal set of light-ray operators in four-dimensional conformal field theories that generate a ${\rm w}_{1+\infty}$ symmetry algebra, and in theories with a continuous global symmetry, we construct a universal set of light-ray operators that generate the $S$ algebra.  The generators of ${\rm w}_{1+\infty}$ are constructed from the stress tensor and are therefore universal to any conformal field theory, while the generators of the $S$ algebra are constructed from the associated conserved current so likewise are universal to conformal field theories with a continuous global symmetry. While parts of our analysis rely on the more tightly constrained framework of conformal field theories, we expect the general structure to extend in some form to generic relativistic theories of quantum fields.  Thus, these too are anticipated to admit a ${\rm w}_{1+\infty}$ symmetry algebra (and $S$-algebra extension in the presence of a continuous global symmetry), although explicit expressions for the generators and/or their algebra may be no longer universal. 

Light-ray operators in Lorentzian conformal field theories have received significant recent attention in the literature  \cite{Caron-Huot:2017vep,Simmons-Duffin:2017nub,Kravchuk:2018htvh,Cordova:2018ygx,Kologlu:2019bco,Kologlu:2019mfz,Caron-Huot:2020adz,Chang:2020qpj,Belin:2020lsr,Besken:2020snx,Kravchuk:2021kwe,Korchemsky:2021htm,Caron-Huot:2022ugt,Chang:2022ryc,Caron-Huot:2022eqs,Hartman:2023qdn,Hartman:2023ccw,Chang:2023szz,Henriksson:2023cnh,Hartman:2024xkw,Homrich:2024nwc,Li:2025knf,Chang:2025zib,Erramilli:2025pfh,Mecaj:2025ecl}
 and a collection of light-ray operators that generate the BMS subalgebra of asymptotic symmetries in asymptotically flat spacetimes was first constructed in the work of \cite{Cordova:2018ygx}.  Our work extends the analysis of \cite{Cordova:2018ygx} and subsequent work of \cite{Besken:2020snx} to the full ${\rm w}_{1+\infty}$ symmetry algebra. A constructive approach to determine the algebra of integrated stress tensors is to compute the commutators directly from the operator product expansion, as in \cite{Besken:2020snx} (see also \cite{Huang:2019fog} for an alternate study of stress-tensor commutators on the lightcone). In \cite{Cordova:2018ygx}, the authors instead rely heavily on the relation of the BMS generators to conserved charges associated to global symmetries of the theory. While this relation no longer holds for all generators of the ${\rm w}_{1+\infty}$ algebra, we nevertheless are able to fix the algebra without directly using the OPE through a judicious application of the Jacobi identity.  

As in \cite{Cordova:2018ygx}, our light-ray operators are not topological, meaning that they do not imply a set of conservation laws.  Indeed, as is well-known in the subject of celestial holography, the associated conservation laws require the presence of propagating gravitons or gauge bosons, since the full conserved charge involves both contributions from the light-ray operators involving the stress-tensor (or current) as well as from soft modes of the graviton (or gauge boson) \cite{Strominger:2017zoo}.    However, as evidenced by more recent work in celestial holography, the existence of \emph{symmetry algebras} can be a powerful tool even if they are not directly connected to conservation laws \cite{Mago:2021wje,Freidel:2021ytz,Ren:2022sws,Guevara:2022qnm,Banerjee:2023zip,Guevara:2024vlc,Guevara:2025tsm}. 

The connection between light-ray operators, asymptotic symmetries, soft theorems, and celestial holography has already appeared in the recent literature \cite{Cordova:2018ygx,He:2019ywq,Donnay:2020fof,Gonzo:2020xza,Freidel:2021ytz,Hu:2022txx,Hu:2023geb,Chang:2025zib,Gonzalez:2025ene,Moult:2025njc}. Our work is a generalization of \cite{Hu:2022txx,Hu:2023geb} (in turn building on \cite{Freidel:2021ytz}) in which the generators of the  ${\rm w}_{1+\infty}$ and $S$ algebras are constructed for free scalar theories in four dimensions. Our expressions reproduce theirs in free theories but strikingly apply to generic \emph{interacting} four-dimensional conformal field theories. Our work, especially our bottom-up approach entirely within conformal field theory, elucidates that the ${\rm w}_{1+\infty}$ symmetry algebra can indeed be understood as a property of genuinely interacting field theories. In this regard, our analysis also provides complementary insight into the asymptotic symmetry analyses in celestial holography, which heavily exploit the approximate free field behavior at null infinity. 

This paper is organized as follows.  We begin with conventions for coordinates, fall-off behavior of fields, and a review of the ${\rm w}_{1+\infty}$ and $S$ algebras in Section \ref{sec:prelim}.  The universal generators and algebra of ${\rm w}_{1+\infty}$ is treated in Section \ref{sec:w}. In Subsections \ref{sec:bms} and \ref{sec:classification}, we exploit a systematic classification of the light-ray operators formed from the stress tensor by scaling dimension and ${\rm SL}(2, \mathbb{C})$ conformal weight to determine explicit expressions for generators of ${\rm w}_{1+\infty}$.  Their algebra is then presented in Subsection \ref{sec:walg} with a brief description of its derivation.  In Section \ref{sec:s}, we carry out the analogous analysis for a spin-one current associated to a continuous global symmetry and find the corresponding $S$ algebra, along with explicit expressions for its generators.  In the penultimate Section \ref{sec:1pt}, we identify a new connection to the physics of asymptotic symmetries by showing that the one-point functions of the ${\rm w}_{1+\infty}$ generators in scalar states are simple re-writings of the universal soft factors in the infinite tower of soft graviton theorems. Finally, in the discussion section \ref{sec:discussion}, we describe additional analyses that will appear in the forthcoming long-form version of this paper \cite{upcoming} and present a number of future research directions that build directly on this work. 

\section{Preliminaries} \label{sec:prelim}

\subsection{Coordinates and Large-$r$ Expansion}

We primarily employ retarded flat Bondi coordinates $(u,r,z,\bar{z})$ \cite{Dumitrescu:2015fej}, which are related to Cartesian coordinates by
\begin{equation} \label{Bondi}
        x^{\mu} = \frac{1}{2}\left(u \partial_z \partial_{\bar{z}} \hat q^{\mu} + r \hat{q}^{\mu}(z,\bar{z})\right),
\end{equation}
with 
\begin{equation}
    \begin{split}
        \hat q^\mu (z, \bar{z}) = \left(1+ z\bar{z}, z+\bar{z}, -i(z-\bar{z}), 1-z \bar{z}\right), \quad \quad 
        \partial_z \partial_{\bar{z}} \hat q^{\mu} = (1,0,0,-1), \quad \quad  \hat q^{\mu}\partial_z \partial_{\bar{z}}\hat{q}_{\mu} = -2,
    \end{split}
\end{equation}
and in which the four-dimensional Minkowski line interval takes the form 
\begin{equation} \label{line-element-urz}
ds^2 = - du dr + r^2 dz d\bar{z}. 
\end{equation} 
We study light-ray operators formed from null integrals of local operators along null infinity $\mathscr{J}$, which is the codimension-one null surface reached by taking the limit $r \to \infty$.  When placed at null infinity, these light-ray operators are known as ``detector'' operators and can be defined in any four-dimensional quantum field theory. In conformal theories, conformal transformations map detector operators at null infinity to light-ray operators in the bulk. We will briefly mention the form of our results under such transformations below, and include full details in  \cite{upcoming}. At null infinity, the stress tensor admits an expansion in large $r$, which is expressed using the following notation
\begin{equation} \label{eq:largergen}
T_{\mu\nu}(u,r,z,\bar{z}) = \sum_{n} \frac{T_{\mu\nu}^{(n)}(u,z,\bar{z})}{r^n}. 
\end{equation}
In conformal field theories, the stress tensor respects the following fall-off conditions
\begin{equation} \label{cft-fall-offs}
    \begin{split}
         T_{uu},T_{uz}, T_{zz} \sim \frac{1}{r^2},\quad \quad T_{z \bar{z}} \sim \frac{1}{r^2}, \quad \quad  T_{rz}, T_{ur} \sim \frac{1}{r^4}, \quad \quad   T_{rr} \sim \frac{1}{r^6},
    \end{split}
\end{equation}
which are justified in more detail in Appendix \ref{sec:conformalsimp}.

\subsection{Asymptotic Symmetry Algebras}

The (wedge subalgebra) of the ${\rm w}_{1 + \infty}$ algebra is an infinite-dimensional algebra generated by an extension of a chiral half of the 4D Poincar\'e algebra. It has appeared recently in the context of celestial holography through the tower of soft theorems that arise in gravitational scattering amplitudes \cite{Guevara:2019ypd,Guevara:2021abz,Strominger:2021lvk,Himwich:2021dau,Freidel:2021ytz,Adamo:2021lrv,Adamo:2021zpw,Himwich:2023njb,Kmec:2024nmu}.  The generators ${\rm w}^{p}_m$ have commutation relations 
\begin{equation} \label{eq:walg}
\left[{\rm w}^{p}_m, {\rm w}^{q}_n\right] = i\left[m(q-1) - n(p-1)\right] {\rm w}^{p+q-2}_{m+n},
\end{equation}
where $p = \frac{3}{2}, 2, \frac{5}{2},3 \ldots,$ and restriction to the range $1-p \leq m \leq p-1$ defines the ``wedge."  There is also a central term with $p=1$, which we will not discuss.\footnote{The next section will demonstrate that, in our context, the presence of a nonzero ${\rm w}^1$ is equivalent to a nontrivial operator appearing in the commutator of two ANEC operators.} In our context, the label $p$ is a chiral ${\rm SL}(2, \mathbb{C})$ weight and $m$ is a chiral ${\rm SL}(2, \mathbb{C})$ mode index. The generators with $p=\frac{3}{2}, 2$ generate a chiral half of the Poincar\'e algebra. The action of generators with $p \geq \frac{5}{2}$ increases the weight, and the full algebra can be finitely generated from the chiral Poincar\'e algebra together with ${\rm w}^{\frac{5}{2}}_{-\frac{3}{2}}$. 

The 4D gauge-theoretic analog of the ${\rm w}_{1+\infty}$ algebra is known as the $S$ algebra \cite{Guevara:2021abz,Strominger:2021lvk,Costello:2022wso,Freidel:2023gue,Kmec:2025ftx}.  The generators $S^{p,a}_m$ have commutation relations  
\begin{equation}
\left[S^{p,a}_m, S^{q,b}_n\right] =  i f^{abc}S^{p+q-1,c}_{m+n}, 
\end{equation}
where $p = 1,\frac{3}{2}, 2, \ldots$ and again the wedge restricts $m$ to lie in the range $1-p \leq m \leq p-1$.  These generators transform in an adjoint representation of ${\rm w}_{1+\infty}$: 
\begin{equation} \label{eq:adjointws}
\left[{\rm w}^{p}_m, S^{q,a}_n\right] = i\left[m(q-1) - n(p-1)\right]S^{p+q-2,a}_{m+n}.
\end{equation}

\section{The ${\rm w}_{1+\infty}$ Algebra of Stress Tensors} \label{sec:w}

In this section, we extend the results of \cite{Cordova:2018ygx,Besken:2020snx} to a larger class of light-ray operators formed from integrals of the stress tensor along a null generator of $\mathscr{J}$.  The organizing principle underlying our analysis is the classification of these operators according to their 4D scaling dimension $\Delta$ and transformation under the Lorentz group ${\rm SO}(1,3)$.  In a conformal field theory, $\Delta$ is the usual eigenvalue under dilations of an operator placed at the origin. To account for transformations under the Lorentz group, we exploit the isomorphism between the 2D conformal group ${\rm SL}(2, \mathbb{C})$ and the Lorentz subgroup ${\rm SO}(1,3)$ of the 4D conformal group and organize the light-ray operators into a basis of operators  that transform as 2D global conformal primary fields under ${\rm SL}(2, \mathbb{C}) \cong {\rm SO}(1,3)$.\footnote{Note this is not the collinear ${\rm SL}(2)$ symmetry discussed in the literature that is formed from a combination of generators of 4D translations, special conformal transformations, Lorentz transformations, and dilations \cite{Braun:2003rp}.} In particular, elements in the basis are thus labeled by left and right Lorentz ${\rm SL}(2,\mathbb{C})$ weights $(h,\bar{h})$ and we will refer to them as ``${\rm SL}(2, \mathbb{C})$ primaries.''\footnote{A similar classification was considered for light-ray operators of free scalars in \cite{Hu:2022txx}. Note that we always use $\Delta$ to refer to the 4D scaling dimension, not to be confused with the 2D boost weight $h+\bar{h}$.}  We then exploit this classification to deduce the algebra generated by these operators.

We proceed systematically, beginning with light-ray operators of the largest scaling dimensions $\Delta$ that can be constructed from the stress tensor. At each fixed scaling dimension, we organize the set into ${\rm SL}(2, \mathbb{C})$ primaries and then determine their commutation relations. Here we summarize the key results of the full analysis in \cite{upcoming}. In this paper we always assume 4D conformal symmetry. 

\subsection{Extended BMS Algebra} \label{sec:bms}

Based on the fall-off conditions \eqref{cft-fall-offs}, there is a single integrated local stress-tensor operator of scaling dimension $\Delta = 1$ and there are no operators with larger scaling dimensions.  This unique, heaviest-weight operator is known as the ``averaged null energy condition operator,'' or ANEC operator for short, and is an ${\rm SL}(2,\mathbb{C})$ primary with the following weights:
\begin{equation}
    \begin{split}
        \mathcal{W}^{\frac{3}{2}}(z, \bar{z}) \equiv \int du~T_{uu}^{(2)}(u, z, \bar{z}), \qquad \left(h,\bar{h}\right) = \left(\frac{3}{2},\frac{3}{2}\right).
    \end{split}
\end{equation}
This operator has been the subject of an extensive literature on positivity constraints and conformal colliders \cite{Hofman:2008ar,Hofman:2016awc,Faulkner:2016mzt,Hartman:2016lgu,Cordova:2017zej,Cordova:2017dhq}; see \cite{Moult:2025nhu} for a recent review and more complete list of references. Notice that the ANEC is a local version of the generator of translations  \eqref{eq:globalP}, which is consistent with its  scaling dimension $\Delta =1$.\footnote{ The action of translations $[P_\mu, \mathcal{O}] \sim -i\partial_\mu \mathcal{O}$ raises the scaling dimension by one.} In Section \ref{sec:1pt}, we will see a precision relation of  $\mathcal{W}^{\frac{3}{2}}$ to the leading soft graviton theorem \cite{Weinberg:1965nx}.

To determine the algebra, we begin by assuming that there are no light neutral scalars with $1 \leq \Delta \leq 2$.  Under this assumption \cite{Besken:2020snx} found, by analyzing the operator product expansion between two stress tensors in conformal field theories in spacetime dimension $D>2$, the identity and the stress tensor are the only operators that contribute to commutators between two stress tensors on the same light sheet. Thus, using their result and neglecting the identity contribution, we conclude that only null integrals of the stress tensor will appear in the commutators that we consider.  It would be interesting to determine the vacuum contributions to our algebra, which intriguingly can depend on the central charges of the theory, but we leave this to a future investigation. 

Since $\mathcal{W}^{\frac{3}{2}}$ is the unique operator of highest scaling dimension $\Delta$, its commutator with itself must vanish,
\begin{equation} \label{eq:ANECcommute}
    \begin{split}
        \left[\mathcal{W}^{\frac{3}{2}}(z, \bar{z}), \mathcal{W}^{\frac{3}{2}}(z', \bar{z}') \right] =0,
    \end{split}
\end{equation}
because there are no operators of scaling dimension $\Delta = 2$ that can be constructed from null integrals of the stress tensor.  This reproduces the commutativity of the ANEC operator found previously by \cite{Casini:2017roe,Cordova:2018ygx,Kologlu:2019bco,Besken:2020snx}.

At the next order in scaling dimension ($\Delta = 0$), the minimal set of independent light-ray operators modulo current conservation \eqref{eq:currentcons} is 
 \begin{equation} \label{subleading-basis}
      \begin{split}
           \int du~ uT_{uu}^{(2)}, \quad \int du~T_{uz}^{(2) },\quad \int du~T_{u\bar{z}}^{(2) }.
     \end{split}
  \end{equation} 
These can be organized into the following ${\rm SL}(2,\mathbb{C})$ primaries: 
\begin{equation}
    \begin{aligned}
        \mathcal{D}(z, \bar{z}) &= \int du~ uT_{uu}^{(2)}(u, z, \bar{z}), &\left(h, \bar{h}\right) = \left(1,1\right), \\     \mathcal{W}^2(z, \bar{z}) &= \frac{1}{2} \int du\left(u \partial_z T_{uu}^{(2)} - 2T_{uz}^{(2)} \right), &\left(h, \bar{h}\right) = \left(2,1\right),
    \end{aligned}
\end{equation}
plus the conjugate of $\mathcal{W}^2$. $\mathcal{D}$ is a local version of the dilation operator \eqref{eq:globalD} and $\mathcal{W}^2$ is a local version of the  generator of Lorentz transformations \eqref{eq:globalJ}.\footnote{Note that the dimension $\Delta = 0$ is consistent with the scaling of the Lorentz action $[J_{\mu \nu}, \mathcal{O}] \sim -i  x_{[\mu} \partial_{\nu]} \mathcal{O}$ and dilation action $[D, \mathcal{O}] \sim -i x^\mu \partial_\mu \mathcal{O}$, which preserve the scaling dimension.} We also note that the $\Delta = 0$ generators are 4D conformal descendants of the ANEC, as observed in \cite{Kologlu:2019mfz,Kologlu:2019bco,Belin:2020lsr}. The 4D descendant structure of these operators will be discussed further in \cite{upcoming}. 

The commutators of the operators we have enumerated thus far with $\Delta = 1$ and $\Delta = 0$ involve contact terms in the transverse directions and have been computed by \cite{Cordova:2018ygx,Besken:2020snx}. They can be smeared in the transverse directions to form a BMS algebra. As discussed explicitly in Section \ref{sec:1pt}, the $\mathcal{W}^2$ generators are related to the subleading soft graviton theorem \cite{Cachazo:2014fwa,Kapec:2014opa}. Here, we simply quote the results for the commutators of $\mathcal{W}^{\frac{3}{2}}$ and $\mathcal{W}^2$, which we will use in the following subsections:
\begin{equation} \label{eq:BMSlocal}
\begin{aligned}
\left[\mathcal{W}^{\frac{3}{2}}(z,\bar{z}), \mathcal{W}^2(z',\bar{z}')\right] &= i \left[ \frac{1}{2} \partial_{z'} -  \partial_z \right]\left(\delta^{(2)}(z-z')\mathcal{W}^{\frac{3}{2}}(z',\bar{z}') \right), \\
\left[\mathcal{W}^2(z,\bar{z}), \mathcal{W}^2(z',\bar{z}')\right] &= i \left[ \partial_{z'} - \partial_z \right]\left(\delta^{(2)}(z-z')\mathcal{W}^{2}(z',\bar{z}') \right). \\
\end{aligned}
\end{equation}

\subsection{Further Light-Ray Operators} \label{sec:classification}

We now continue our general classification of null-integrated stress tensor operators, organized by dilation weight  $\Delta$ and Lorentz ${\rm SL}(2,\mathbb{C})$ primary structure, to those with $\Delta < 0$. At $\Delta = -1$, we find a minimal set of light-ray operators, again modulo current conservation, given by
\begin{equation} \label{eq:SubSubleadingSet}
\int du~ u^2 T_{uu}^{(2)}, \ \  \int du~  u T_{uz}^{(2)},    \ \  \int du~   T_{zz}^{(2)}, \ \ \int du~   T_{z\bar{z}}^{(2)}, 
\end{equation}
and their conjugates. Next, organizing \eqref{eq:SubSubleadingSet} and their $z, \bar{z}$ derivatives into ${\rm SL}(2,\mathbb{C})$ primaries with increasing left-weight $h$ (and their conjugates), we find 
\begin{equation}
\begin{aligned}
\mathcal{L}_0(z,\bar{z}) &\equiv \int du~ u^2 T_{uu}^{(2)}, \qquad &\left(h, \bar{h}\right) = \left(\frac{1}{2},\frac{1}{2}\right), \\
\mathcal{X}^{\frac{3}{2}}(z,\bar{z}) &\equiv \int du \left(u^2 \partial_z T_{uu}^{(2)} -  u T_{uz}^{(2)}  \right), \qquad &\left(h, \bar{h}\right) = \left(\frac{3}{2},\frac{1}{2}\right), \\
\mathcal{K}(z,\bar{z}) &\equiv \int du \left(u^2 \partial_z\partial_{\bar{z}}T_{uu}^{(2)} + T_{z\bar{z}}^{(2)} - u \partial_z T_{u\bar{z}}^{(2)} - u \partial_{\bar{z}}T_{uz}^{(2)} \right), \qquad &\left(h, \bar{h}\right) = \left(\frac{3}{2},\frac{3}{2}\right), \\
\mathcal{W}^{\frac{5}{2}}(z,\bar{z}) &\equiv \frac{1}{8}\int du\left(  u^2 \partial_{z}^2 T_{uu}^{(2)} - 4 u \partial_z T_{uz}^{(2)} + 6 T_{zz}^{(2)}  \right), \qquad  &\left(h, \bar{h}\right) = \left(\frac{5}{2},\frac{1}{2}\right). 
\end{aligned}
\end{equation} 
The reasoning behind this choice of labeling is as follows. The notation $\mathcal{L}_0$ is chosen to match the notation of \cite{Casini:2017roe,Besken:2020snx,Belin:2020lsr,Huang:2020ycs,Huang:2021hye}, where a Virasoro algebra was discussed between operators of the form $\mathcal{L}_{n} = \int du~ u^{n+2} T_{uu}^{(2)}$. Such operators appear at further lower orders in $\Delta$ and  are described in more depth in \cite{upcoming}. The notation $\mathcal{K}$ is chosen because it is a local version of the generator of 4D special conformal transformations $K$ \eqref{eq:globalK}. The normalization of $\mathcal{W}^{\frac{5}{2}}$ is chosen in anticipation of the ${\rm w}_{1+\infty}$ structure we will see shortly. An explicit relation of $\mathcal{W}^{\frac{5}{2}}$ to the subsubleading soft graviton theorem is presented in Section \ref{sec:1pt}. Stress tensor expressions for the matter contribution to the ``hard charge'' associated to the subsubleading soft graviton theorem that are similar to our formula for $\mathcal{W}^{\frac{5}{2}}$ have appeared previously in the literature \cite{Campiglia:2016jdj,Campiglia:2016efb,Horn:2022acq}. We also note that $\mathcal{W}^{\frac{5}{2}}$ is a 4D descendant of $\mathcal{D}$ and $\mathcal{W}^{2}$ and will discuss the 4D descendant structure of the full collection of the $\mathcal{W}^p$ generators further in \cite{upcoming}.  

For the purposes of this summary paper, we focus on the commutators of the $\mathcal{W}^p$ generators. The commutators of $\mathcal{W}^p$ for $p=\frac{3}{2},2,\frac{5}{2}$ are fully specified by \eqref{eq:ANECcommute} and \eqref{eq:BMSlocal} plus the additional commutators 
\begin{equation} \label{eq:52examples}
\begin{aligned}
\left[\mathcal{W}^{\frac{3}{2}}(z,\bar{z}), \mathcal{W}^\frac{5}{2}(z',\bar{z}')\right] &= i \left[ \frac{1}{2} \partial_{z'} -  \frac{3}{2}\partial_z \right]\left(\delta^{(2)}(z-z')\mathcal{W}^{2}(z',\bar{z}') \right) + i \frac{3}{4} \partial_z^2\left(\delta^{(2)}(z-z')\mathcal{D}(z',\bar{z}') \right),  \\
\left[\mathcal{W}^2(z,\bar{z}), \mathcal{W}^\frac{5}{2}(z',\bar{z}')\right] &= i \left[ \partial_{z'} - \frac{3}{2}\partial_z \right]\left(\delta^{(2)}(z-z')\mathcal{W}^{\frac{5}{2}}(z',\bar{z}') \right) + i \frac{3}{16} \partial_z^3\left(\delta^{(2)}(z-z')\mathcal{L}_0(z',\bar{z}') \right). \\
\end{aligned}
\end{equation}
The derivation of \eqref{eq:52examples} including the necessary assumptions will be detailed in \cite{upcoming}. We will see shortly that the rightmost terms in each of the above commutators do not contribute to the wedge. 

The next order $\Delta = -2$ exhibits a more intricate structure, which generalizes to further lower scaling dimensions. Using current conservation, the minimal set is
\begin{equation} \label{eq:Sub3leadingSet}
\begin{aligned}
&\int du~ u^3 T_{uu}^{(2)}, \ \  \int du~  u^2 T_{uz}^{(2)}, \ \   \  \int du~   uT_{zz}^{(2)}, \ \  \  \int du~   u T_{z\bar{z}}^{(2)},\  \  \ \int du~ T_{zz}^{(3)}, \ \ \ \int du~ T_{rz}^{(4)}, 
\end{aligned}
\end{equation}
and their conjugates. We again organize these into ${\rm SL}(2,\mathbb{C})$ primaries with increasing left-weight $h$, which will be explicitly presented in \cite{upcoming}. The lowest weight primary is $\mathcal{L}_{1} \equiv \int du~ u^3 T_{uu}$ with weights $(h,\bar{h}) = (0,0)$. For the purpose of deriving the ${\rm w}_{1+\infty}$ algebra, we focus on primaries of left-weight $h=3$. At $h=3$ there is no local primary of weight $(3,0)$, which is perhaps not surprising since the stress tensor is spin two.  However, there is a local primary of weight $(h, \bar{h}) = (3, 1)$:
\begin{equation} \label{eq:W3des}
\begin{aligned}
\partial_{\bar{z}}\mathcal{W}^{3}(z,\bar{z})  = 
  \frac{1}{48}\int du \left( u^3 \partial_z^3 \partial_{\bar{z}} T_{uu}^{(2)} - 6 u^2 \partial_z^2 \partial_{\bar{z}} T_{uz}^{(2)} + 18 u \partial_z \partial_{\bar{z}} T_{zz}^{(2)} + 24\left( T_{zz}^{(3)} + \partial_z T_{rz}^{(4)} \right) \right).  \end{aligned}
\end{equation}
Here we use the notation $\partial_{\bar{z}}\mathcal{W}^{3}$ to indicate that this is an ${\rm SL}(2,\mathbb{C})$ right-primary descendant that descends from a primary $\mathcal{W}^{3}$
of weight $(3,0)$. In a 4D CFT, there is no restriction on the sign of ${\rm SL}(2,\mathbb{C})$ weights $(h,\bar{h})$ of the light-ray operators. Generally, if an ${\rm SL}(2,\mathbb{C})$ primary has a non-positive weight $h$, it has a primary descendant of weight $1-h$ (respectively $\bar{h}$ and $1-\bar{h}$). We always use ``primary descendant" to refer to an ${\rm SL}(2,\mathbb{C})$ primary descendant, not to be confused with the 4D primary descendants discussed e.g. in \cite{Chang:2020qpj}.

Although $\mathcal{W}^{3}$ has a local (in $z,\bar{z}$) primary descendant, it appears a priori non-local itself. However, we observe that in the case of a conformally coupled free scalar,\footnote{In our conventions $T_{\mu\nu} = \nabla_{\mu}\phi \nabla_{\nu}\phi - \frac{1}{4(d-1)}\left((d-2) \nabla_{\mu}\nabla_{\nu} + \eta_{\mu\nu}\nabla_{\rho}\nabla^{\rho}\right) \phi^2$ 
in flat space and we use the large-$r$ expansion of scalar wave equation $(n-1)\partial_u \phi^{(n)} + \partial_z \partial_{\bar{z}} \phi^{(n-1)} = 0$. Note that free scalar field theory violates our assumptions of no light neutral scalars.  However, for the free scalar case, the deviations from the structure we present here are fairly mild and under control, and will be discussed in detail in \cite{upcoming}.} the apparently non-local contribution can be rewritten as 
\begin{equation}
  24 \left(T_{zz}^{(3)} + \partial_z T_{rz}^{(4)}\right) = 12 \int du'~ {\rm sgn}(u-u')~ \partial_{\bar{z}} \left( \partial_z^2\phi^{(1)}(u)\partial_z\phi^{(1)}(u') \right),
\end{equation}
where ${\rm sgn}(u) = 2 \int_{-\infty}^{\infty} \frac{d\omega}{2\pi i} \frac{1}{\omega} e^{i \omega u}$. Thus, we can strip off a $\partial_{\bar{z}}$ derivative from \eqref{eq:W3des} and find a local expression in the transverse directions $(z,\bar{z})$ at the price of introducing non-locality in $u$. For free scalars, this pattern continues at further subleading orders, and it would be interesting to understand the extent to which this phenomenon persists in generic interacting theories, for instance $\mathcal{N}=4$ super-Yang-Mills. 

We now proceed to general  $\Delta \leq -2 $ and introduce $\beta = 1-\Delta \geq 3$.  At each $\beta$, a minimal set is
\begin{equation} \label{eq:SubnleadingSet}
\begin{aligned}
&\int du~ u^{\beta} T_{uu}^{(2)}, \ \  \int du~  u^{\beta-1} T_{uz}^{(2)}, \ \   \  \int du~   u^{\beta - n_1} T_{zz}^{(n_1)},  \ \ \  \int du~   u^{\beta - n_1} T_{z\bar{z}}^{(n_1)}, \ \ n_1 \in [2,\beta], \\ 
& \qquad \int du~ u^{\beta - n_2}T_{rz}^{(n_2+1)}, \ \  n_2 \in [3,\beta],  \ \ \  \int du~ u^{\beta-n_3}T_{rr}^{(n_3+2)}, \ \ \ n_3 \in [4,\beta],
\end{aligned}
\end{equation}
and their conjugates. Again, at every order, we organize this set into ${\rm SL}(2,\mathbb{C})$ primaries by increasing left-weight $h$, with the lowest-weight primary being $\mathcal{L}_{\beta-2} \equiv\int du~ u^{\beta} T_{uu}^{(2)}$ of weight $(h,\bar{h}) = \left(\frac{3-\beta}{2},\frac{3-\beta}{2}\right)$. For the purpose of studying the $\rm{w}_{1+\infty}$ algebra, we are interested in the highest-left-weight $h = \frac{3+\beta}{2}$ primary. The explicit enumeration of all of the primaries and primary descendants at general $\Delta$ will be presented in \cite{upcoming}, although a general study of their commutators (all of which we expect to be highly constrained) will be left to future work.

As in the case of $\Delta = -2$, we find a local primary descendant of weight $\left(h, \bar{h}\right) = \left(\frac{3+\beta}{2}, \frac{\beta-1}{2}\right)$:  
\begin{equation} \label{eq:wdes}
  \begin{aligned}
\partial_{\bar{z}}^{\beta-2}&\mathcal{W}^{\frac{3+\beta}{2}}(z,\bar{z}) \\ = 
    &\frac{1}{2^{\beta} \beta!}\int du~ \Bigg( \left(u \partial_z \partial_{\bar{z}} \right)^{\beta-2} \left[u^2 \partial_z^{2} T_{uu}^{(2)} - 2 \beta u\partial_z T_{uz}^{(2)} + 3 \beta(\beta-1)T_{zz}^{(2)} \right]  \\
    &\quad + \sum_{n=3}^{\beta} \frac{\beta!(n-3)!}{(\beta-n)!} \left(u\partial_z \partial_{\bar{z}}\right)^{\beta-n} \left[(n{+}1)(n{-}2) T_{zz}^{(n)} + 2(n{-}1)\partial_z T_{rz}^{(n+1)} + \partial_z^2 T_{rr}^{(n+2)} \right] \Bigg),
  \end{aligned}
\end{equation}
which descends from a primary $\mathcal{W}^{\frac{3+\beta}{2}}$ with weight $\left(h, \bar{h}\right) = \left(\frac{3+\beta}{2}, \frac{3-\beta}{2}\right)$. To compare with \eqref{eq:W3des}, recall that $T_{rr}^{(5)} = 0$ \eqref{cft-fall-offs}. For free scalars, at every order, the nonlocality in the transverse directions can be traded for nonlocality in $u$. We expect that at all orders (and have explicitly checked at several low orders) the non-local-in-$u$ free scalar expressions exactly match those appearing in \cite{Hu:2022txx,Hu:2023geb}. In Section \ref{sec:1pt}, we find that the one-point functions of \eqref{eq:wdes} also involve $\beta-2$ derivatives in $\bar{z}$ that can be stripped off, leading to a direct relation between the expectation values of the primaries $\mathcal{W}^{\frac{3+\beta}{2}}$ and soft factors associated to sub${}^\beta$-leading soft graviton theorems.  

In coordinates  $y^- = u$, $y^+ = -\frac{1}{r}$, the leading (conformally rescaled) term in the expansion of the stress tensor is simply $T_{\mu\nu}(y^+ = 0)$ and the subleading large-$r$ expansion in the above expressions becomes a derivative expansion in $y^+$. This will be discussed further in \cite{upcoming}. 

\subsection{${\rm w}_{1+\infty}$ Algebra} \label{sec:walg}

In this section, we sketch the proof, presented in detail in \cite{upcoming}, that modes of  the  $\mathcal{W}$ operators defined in the previous subsection obey the ${\rm w}_{1+\infty}$ algebra. To do so, we switch to the conventional labels $p = \frac{3+\beta}{2}$. In this notation, $p = \frac{3}{2}, 2, \frac{5}{2}, 3, \ldots$, and  
$\mathcal{W}^{p}(z,\bar{z})$ has $\Delta = 4 - 2p$ with $(h,\bar{h}) = (p, 3-p)$. The main result is that the $\mathcal{W}^p$ generators satisfy the local algebra
\begin{equation} \label{eq:localW}
\left[\mathcal{W}^p(z,\bar{z}), \mathcal{W}^q(z',\bar{z}')\right] = i \left[ (p-1) \partial_{z'} - (q-1) \partial_z \right]\left(\delta^{(2)}(z-z')\mathcal{W}^{p+q-2}(z',\bar{z}') \right)+ \cdots
\end{equation}
where contributions from the identity are neglected and the dots denote terms that vanish outside the wedge. Namely, these terms all vanish upon considering the modes 
\begin{equation}
{\rm w}^{p}_{m} \equiv \int  d^2z~ z^{p+ m - 1} \mathcal{W}^{p}(z,\bar{z}), \qquad  1-p \leq m \leq p-1.
\end{equation}
Integrating \eqref{eq:localW} thus gives the wedge subalgebra of the ${\rm w}_{1+\infty}$ algebra, repeated here for convenience: 
\begin{equation}
\left[{\rm w}^{p}_{m}, {\rm w}^{q}_{n} \right] = i \left[m(q-1) - n(p-1)\right] {\rm w}^{p+q-2}_{m+n}. 
\end{equation}

The proof of \eqref{eq:localW} proceeds by induction in $p+q$. The base case $p+q = 4$ corresponds to commutators $\left[\mathcal{W}^{\frac{3}{2}}(z,\bar{z}), \mathcal{W}^{\frac{5}{2}}(z',\bar{z}')\right]$ and $\left[\mathcal{W}^2(z,\bar{z}),\mathcal W^{2}(z',\bar{z}')\right]$, which are explicitly demonstrated to take the form \eqref{eq:localW}, specifically \eqref{eq:BMSlocal} and \eqref{eq:52examples}. Then, given our explicit form of $\mathcal{W}^p$, one can compute the action of the global translation generator $P_u$ (see \eqref{eq:Paction}):
\begin{equation}
\left[P_u, \mathcal{W}^p(z,\bar{z})\right] = \frac{i}{2} \partial_z\mathcal{W}^{p-\frac{1}{2}}(z,\bar{z}).
\end{equation}
Next, assuming that \eqref{eq:localW} holds for $p+q=n-\frac{1}{2}$, acting with the translation $P_u$ on a commutator with $p+q = n$ and using the Jacobi identity implies 
\begin{equation}
\left[P_u, \left[\mathcal{W}^p(z,\bar{z}), \mathcal{W}^q(z',\bar{z}')\right] \right] = \frac{i}{2}\left( \partial_z\left[\mathcal{W}^{p-\frac{1}{2}}(z,\bar{z}), \mathcal{W}^q(z',\bar{z}')\right] + \partial_{z'}\left[\mathcal{W}^p(z,\bar{z}), \mathcal{W}^{q-\frac{1}{2}}(z',\bar{z}')\right]\right) .
\end{equation}
The inductive assumption then constrains the right-hand-side of the commutator with $p+q=n$ to take the form \eqref{eq:localW} up to terms that are in the kernel of $P_u$. At every order there are a small number of such terms that could appear ``inside the wedge." The presence of these terms can be considered case-by-case and shown to be in contradiction with the inductive assumption, which completes the proof.   

\section{The $S$ algebra of Spin-one Currents} \label{sec:s}

In this section, we perform the same analysis for spin-one non-abelian global symmetry currents as we did for the stress tensor in the previous section.  We denote the spin-one conserved current by $j_{\mu}^{a}$, where $a$ is the adjoint index of the associated symmetry group $G$. Again, we proceed systematically by using the falloff conditions \eqref{eq:currentfalloffs} to enumerate the operators of decreasing scaling dimension $\Delta$  and then organizing them into ${\rm SL}(2,\mathbb{C})$ primaries. In this summary paper, we present only the primaries relevant for the $S$ algebra, and  further details on the other primaries will be presented in \cite{upcoming}. In deriving the algebra, as in the previous section, we assume that only null integrals of the current can appear in the commutator between two null-integrated currents. 

\subsection{Leading Non-abelian Charge Algebra}

The leading scaling dimension light-ray operator built from $j_{\mu}^a$ has $\Delta = 0$,  was considered in \cite{Hofman:2008ar,Beane:2015ufo,Cordova:2018ygx,Kologlu:2019mfz}, and is given by
\begin{equation}
  \mathcal{S}^{1,a}(z,\bar{z}) = \int du~ j_u^{a \, (2)}, \qquad (h, \bar{h}) = \left(1,1\right).
\end{equation}
This is an ${\rm SL}(2,\mathbb{C})$ primary and as identified in \cite{Cordova:2018ygx}, the commutator of two $\mathcal{S}^{1,a}$ generators is constrained to take the form
\begin{equation} \label{local-leading}
\left[\mathcal{S}^{1,a}(z,\bar{z}), \mathcal{S}^{1,b}(z',\bar{z}')\right] = i f^{abc} \delta^{(2)}(z-z') \mathcal{S}^{1,c}(z',\bar{z}'),
\end{equation}
where $f^{abc}$ are the structure constants of $G$. Then, integrating over the transverse coordinates
\begin{equation}
    Q^{a} = \int d^2z~ \mathcal{S}^{1,a}(z,\bar{z}),
\end{equation}
the charges $Q^a$ obey the algebra 
\begin{equation}
\left[Q^a, Q^b\right] = i f^{abc} Q^c. 
\end{equation}
More generally, the commutation relation \eqref{local-leading} can be smeared against arbitrary functions to reproduce the asymptotic symmetry algebra in non-abelian gauge theory that is related to the leading soft gluon theorem \cite{Weinberg:1965nx,Strominger:2013lka}.  

\subsection{Further Light-ray Operators and the $S$ algebra}

We now proceed to smaller $\Delta$. First, at $\Delta = -1$, we have a minimal set of light-ray operators
\begin{equation}
\int du~ u j_u^{a \, (2)}, \qquad \int du~ j_z^{a \,(2)}, \qquad \int du~ j_{\bar{z}}^{a \, (2)}, 
\end{equation}
modulo current conservation \eqref{eq:jcons}. The associated primaries are 
\begin{equation}
\begin{aligned}
\mathcal{J}^{\frac{1}{2},a}(z,\bar{z}) &= \int du~ u j_u^{a \, (2)}, \qquad &\left(h,\bar{h}\right) = \left(\frac{1}{2}, \frac{1}{2} \right), \\
\mathcal{S}^{\frac{3}{2},a} (z,\bar{z}) &= \frac{1}{2}\int du \left(u \partial_z j_u^{a \,(2)} - j_{z}^{a \, (2)} \right) \qquad 
&\left(h,\bar{h}\right) = \left( \frac{3}{2},  \frac{1}{2}\right),
\end{aligned}
\end{equation}
and the complex conjugate of $\mathcal{S}^{\frac{3}{2},a}$. The operator $\mathcal{S}^{\frac{3}{2},a}$ is related to the subleading soft theorem in gauge theory \cite{GellMann1954,Low1954,Low1958,Burnett1968} and the expression in terms of currents appears  for the QED case first in \cite{Lysov:2014csa}. Both $\mathcal{S}^{\frac{3}{2},a}$ and $\mathcal{J}^{\frac{1}{2},a}$ are 4D conformal descendants of the leading operator $\mathcal{S}^{1,a}$. The 4D descendant structure of the $S$-algebra generators will be discussed further in \cite{upcoming}.  

At $\Delta = -2$, a minimal set modulo current conservation is 
\begin{equation}
\int du~ u^2 j_u^{a \, (2)}, \qquad \int du~ u j_z^{a \,(2)}, \qquad \int du~ j_{z}^{a \, (3)}, \qquad \int du~  j_{r}^{a \, (4)},
\end{equation}
and their conjugates. Again, we organize these into primaries with increasing left-weight $h$, starting with the lowest-weight operator $\mathcal{J}^{0,a}(z,\bar{z}) \equiv \int du~ u^2 j_u^{a \, (2)}$ with $(h,\bar{h}) = (0,0)$. To study the $S$ algebra, we focus on primaries with left-weight $h=2$, where we see a structure of primary descendants in analogy with the stress tensor case. Here, due to the fact that the current is a spin-one operator, there is no local primary of weight $(2,0)$.  However, there is a local spin-one primary descendant
\begin{equation} \partial_{\bar{z}}\mathcal{S}^{2,a}(z,\bar{z}) = \frac{1}{8}\int du~ \left( u^2 \partial_z^2 \partial_{\bar{z}} j_u^{(2)} - 2 u \partial_z\partial_{\bar{z}} j_{z}^{(2)} - 2 \partial_z j_r^{(4)} - 2 j_z^{(3)} \right), \qquad (h,\bar{h}) = (2,1).  
\end{equation}

The same structure continues at general scaling dimension. At general $\Delta \leq -2$, let $\alpha = - \Delta\geq 2$. A minimal set of light-ray operators modulo current conservation is given by
\begin{equation}
\int du~ u^{\alpha} j_u^{a \, (2)}, \quad \int du~ u^{\alpha - m_1} j_z^{a \,(m_1 + 1 )}, \ m_1 \in [1,\alpha], \quad \int du~  u^{\alpha-m_2}j_{r}^{a \, (m_2 + 2)}, \ m_2 \in [2,\alpha],
\end{equation}
and their conjugates.  Again we identify ${\rm SL}(2,\mathbb{C})$ primaries with increasing left-weight $h$, beginning with the lowest-weight primary
$\mathcal{J}^{\frac{2-\alpha}{2},a}(z,\bar{z}) \equiv  \int du~ u^{\alpha} j_u^{a \, (2)}$ with weight $(h,\bar{h}) = \left(\frac{2-\alpha}{2},\frac{2-\alpha}{2}\right)$. For the purpose of the $S$ algebra, we focus on the generators with highest left weight $h = \frac{2+\alpha}{2}$. Here we find a primary descendant with weights $\left(\frac{2+\alpha}{2}, \frac{\alpha}{2}\right)$:
\begin{equation} \label{eq:Sdes}
\begin{aligned}
\partial_{\bar{z}}^{\alpha -1}\mathcal{S}^{\frac{2+\alpha}{2}}(z, \bar{z}) &= \frac{1}{2^{\alpha}\alpha!}\int du \Bigg(\left(u \partial_z \partial_{\bar{z}}\right)^{\alpha-1} \left[u\partial_z j_u^{(2)} - \alpha j_z^{(2)} \right] \\ &\qquad \qquad \qquad ~ - \sum_{m=2}^{\alpha} \frac{\alpha!(m-2)!}{(\alpha-m)!}\left(u \partial_z \partial_{\bar{z}}\right)^{\alpha-m} \left[(m{-}1) j_z^{(m+1)} + \partial_z j_r^{(m+2)} \right] \Bigg),
\end{aligned}
\end{equation}
which descends from a primary $\mathcal{S}^{\frac{2+\alpha}{2}}(z, \bar{z})$ with weights $\left(\frac{2+\alpha}{2}, \frac{2-\alpha}{2}\right)$. These operators are related to a tower of subleading soft gluon theorems, and an expression for (a right-descendant of) $\partial_{\bar{z}}^{\alpha -1}\mathcal{S}^{\frac{2+\alpha}{2}}(z, \bar{z})$ appears in \cite{Campiglia:2018dyi}. In analogy with the previous section, in the free scalar construction of these generators, the non-locality in the transverse directions can be explicitly traded for non-locality in $u$. We again expect that at all orders the non-local-in-$u$ free scalar expressions exactly match those appearing explicitly in \cite{Hu:2023geb}. 

To derive the $S$ algebra, we  switch to the notation $p = \frac{2+\alpha}{2}$. In this notation the operators $\mathcal{S}^{p,a}$ have $p=1,\frac{3}{2},2,\dots$, scaling dimensions $\Delta = 2-2p$, and ${\rm SL}(2,\mathbb{C})$ weights $(p, 2-p)$. 
As in the proof of the ${\rm w}_{1+\infty}$ algebra presented in Section \ref{sec:walg}, we can prove by induction in $p+q$ that the local commutator of the $\mathcal{S}^{p,a}$ operators takes the form 
\begin{equation} \label{eq:localS}
\left[\mathcal{S}^{p,a}(z,\bar{z}), \mathcal{S}^{q,b}(z',\bar{z}')\right] = i f^{abc} \delta^{(2)}(z-z') \mathcal{S}^{p+q-1,c}(z',\bar{z}') + \cdots, 
\end{equation}
where the dots vanish upon  restriction to the wedge. Namely, these terms all vanish upon considering the modes 
\begin{equation}
{S}^{p,a}_{m} \equiv \int  d^2z~ z^{p+ m - 1} \mathcal{S}^{p,a}(z,\bar{z}), \qquad  1-p \leq m \leq p-1, 
\end{equation}
and integrating \eqref{eq:localS} gives the wedge subalgebra of the $S$ algebra, repeated here for convenience:
\begin{equation}
\left[S^{p,a}_m, S^{q,b}_n\right] = i f^{abc} S^{p+q-1,c}_{m+n}. 
\end{equation}
The induction proof again uses the Jacobi identity with $P_u$, which has the explicit action
\begin{equation}
\left[P_u, \mathcal{S}^{p,a}(z,\bar{z}) \right] = \frac{i}{2} \partial_z S^{p-\frac{1}{2},a}(z,\bar{z}). 
\end{equation}
The  proof proceeds in direct analogy with the stress-tensor case in Section \ref{sec:walg} and the details will be presented in \cite{upcoming}.

\section{One-point Functions and Soft Factors} \label{sec:1pt}

The class of light-ray operators that is studied in this paper was largely inspired by investigations in celestial holography. In particular, we expect that they can be identified with the matter contributions to the ``hard'' charges that generate the asymptotic ${\rm w}_{1+\infty}$ and $S$-symmetry algebras.\footnote{In \cite{Hu:2022txx} (see also \cite{Freidel:2021ytz}), it was shown using canonical commutation relations that the expressions for $\mathcal{W}^p$ \eqref{eq:wdes}, specialized to the case of free fields at null infinity, generate the ${\rm w}_{1+\infty}$ symmetry  action on massless particles found in \cite{Himwich:2021dau} and derived directly from soft factors in \cite{Himwich:2023njb}.} In this section, we present a \emph{new},  although closely related, precise connection to the physics of asymptotically flat spacetimes.  Specifically, we show that the one-point functions of the ${\rm w}_{1+\infty}$ light-ray operators evaluated in momentum eigenstates of scalar fields can be identified with (derivatives of) the universal soft factors that arise in low-energy limits of scattering amplitudes.  Intriguingly, this precise connection does not extend to one-point functions between spinning momentum eigenstates, but perhaps this mismatch can be exploited in a clever way to find new constraints in conformal field theory or quantum gravity in asymptotically flat spacetimes.\footnote{For example, the one-point function between stress-energy tensors famously depends on the $a$ and $c$ anomaly coefficients \cite{Duff:1977ay,Deser:1976yx,CARDY1988749,OSBORN198997,Komargodski:2011vj}.  In contrast, there is no corresponding pair of free parameters in the soft factors, especially at leading orders, even for spinning particles.} 

The starting point for the derivation of one-point functions between scalar operator states is the three-point conformal correlation function for a stress tensor and two scalar operators of scaling dimension $\Delta$, given by \cite{Osborn:1993cr}
\begin{equation} \label{tOO-3point}
    \begin{split}
        \langle \mathcal{O}(x_1) T_{\mu\nu}(x_2) \mathcal{O} (x_3)\rangle
         = \frac{a}{x_{12}^4 x_{23}^4 x_{13}^{2\Delta-4}} \left(\frac{X_{13}{}_\mu X_{13}{}_\nu}{X_{13}^2}- \frac{1}{4} \eta_{\mu\nu}\right), 
    \end{split}
\end{equation}
where
\begin{equation}
    \begin{split}
        X_{13}^\mu \equiv \frac{x_{12}^\mu}{x_{12}^2} + \frac{x_{23}^\mu}{x_{23}^2}.
    \end{split}
\end{equation}
This correlation function is entirely fixed by conformal symmetry and the coefficient $a$ is related to the normalization of the scalar two-point function by the stress-tensor Ward identity.  We will use the Lorentzian signature correlator in which the operators are ordered as written and this ordering is implemented by the $i \epsilon$ prescription $x_{ij}^2 \to -(t_{ij}-i \epsilon)^2 + |\vec{x}_{ij}|^2$ for $i<j$. Setting the normalization of the two-point function to be unity
\begin{equation} \label{2-point}
    \begin{split}
        \langle \mathcal{O}(x_1) \mathcal{O}(x_2) \rangle = \frac{1}{\left( -(t_{12}-i \epsilon)^2 + |\vec{x}_{12}|^2 \right)^\Delta}, 
    \end{split}
\end{equation}
then in four dimensions, 
\begin{equation}
    \begin{split}
        a = - \frac{2 \Delta}{3 \pi^2}. 
    \end{split}
\end{equation}

The one-point correlation function of the ANEC can be derived from \eqref{tOO-3point} by contour integration \cite{Hofman:2008ar,Cordova:2018ygx,Belin:2020lsr} 
\begin{equation}
    \begin{split}
        \langle \mathcal{O}(x_1) \mathcal{W}^{\frac{3}{2}}(z , \bar{z}) \mathcal{O} (x_3)\rangle=
        \langle \mathcal{O}(x_1)\int du~ T_{uu}^{(2)}(u  ,  z , \bar{z}) \mathcal{O} (x_3)\rangle
         = \frac{3a \pi i}{ x_{13}^{2\Delta-2}(x_{13} \cdot \hat q+i\epsilon) ^3}.
    \end{split}
\end{equation}
The relation to the universal soft factors in soft theorems is made explicit by Fourier-transforming the positions of the two scalar operators.  Specifically, we find
\begin{equation}
    \begin{split}
       \langle \mathcal{O}(p_1)|\mathcal{W}^{\frac{3}{2}} ( z,\bar{z})|\mathcal{O} (p_3)\rangle
       &\equiv\int d^4 x_1 \int d^4 x_3 ~e^{-ip_1\cdot x_1+i p_3 \cdot x_3}\langle \mathcal{O}(x_1) \mathcal{W}^{\frac{3}{2}}(z , \bar{z}) \mathcal{O} (x_3)\rangle
       \\&= \frac{1}{4 \pi }\frac{(-p_1^2)^2}{(-  p_1 \cdot \hat n)^3} \langle \mathcal{O}(p_1)|\mathcal{O}(p_3)\rangle,
    \end{split}
\end{equation}
where in our choice of coordinates \eqref{Bondi} $\hat n^\mu \equiv \frac{1}{2} \hat q^\mu$ is the properly normalized ``unit" null vector.  We have written our final expression in terms of the Fourier transform of the two-point function:
\begin{equation} \label{eq:twopt}
    \begin{split}
        \langle \mathcal{O}(p_1)|\mathcal{O}(p_2)\rangle &\equiv \int d^4 x_1 \int d^4 x_2 ~e^{-ip_1\cdot x_1+i p_2 \cdot x_2}\langle \mathcal{O}(x_1) \mathcal{O}(x_2) \rangle\\
       & =(2 \pi)^4 \delta^{(4)}(p_1-p_2)\frac{(2 \pi)^3 \Theta(p_1^0)}{ \Gamma(\Delta) \Gamma(\Delta-1) }\frac{(-p_1^2)^{\Delta-2}}{2^{2\Delta-2}}. 
    \end{split}
\end{equation}
To calculate the one-point functions for the other generators of the ${\rm w}_{1+\infty}$ algebra, we relate the result to the one-point function for the ANEC operator.  Specifically,  
\begin{equation}
    \begin{split}
        \langle\mathcal{O}(x_1) \mathcal{W}^{2}&(z,\bar{z}) \mathcal{O}(x_3)\rangle\\
        &=  -\frac{1}{4} \left[   3 (\partial_z \hat q \cdot x_1) +( \hat q \cdot x_1)\partial_z  +  3 (\partial_z \hat q \cdot x_3) +( \hat q \cdot x_3)\partial_z \right] 
        \langle \mathcal{O}(x_1) \mathcal{W}^{\frac{3}{2}}(z,\bar{z}) \mathcal{O}(x_3) \rangle,
    \end{split}
\end{equation}
and for $\beta\geq 2$ up to and including $\beta =5$,  we find by explicit calculation that 
\begin{equation}
    \begin{split}
        \langle\mathcal{O}&(x_1) \partial_{\bar{z}}^{\beta-2}\mathcal{W}^{\frac{3+\beta}{2}}(z,\bar{z}) \mathcal{O}(x_3)\rangle\\
        &= \partial_{\bar{z}}^{\beta-2} \Bigg(\frac{(-1)^\beta}{2^{\beta+1} \beta!} \left[ \prod_{m=3}^{\beta+2}\left( m (\partial_z \hat q \cdot x_1) +( \hat q \cdot x_1)\partial_z\right) +\prod_{m=3}^{\beta+2}\left( m (\partial_z \hat q \cdot x_3) +( \hat q \cdot x_3)\partial_z\right)\right]\\& \quad \quad\quad \quad\quad \quad \quad \quad\quad \quad\quad \quad\quad \quad\quad \quad\quad \quad\quad \quad\quad \quad\quad \quad\quad \quad  \times 
        \langle \mathcal{O}(x_1) \mathcal{W}^{\frac{3}{2}}(z,\bar{z}) \mathcal{O}(x_3)\rangle\Bigg),
    \end{split}
\end{equation}
which we expect to extend to all larger values of $\beta$.  Note that unlike the generalized light-ray operators of the stress tensor $\mathcal{L}_n$ studied in \cite{Casini:2017roe,Cordova:2018ygx,Besken:2020snx,Belin:2020lsr}, the one-point functions of our ${\rm w}_{1+\infty}$ generators appear to be finite.\footnote{We have explicitly checked this for the $\beta =5$ generator, which involves $\int du ~u^5 \partial_z^5 T_{uu}^{(2)}$, since the authors of \cite{Cordova:2018ygx,Belin:2020lsr} identified the operator $\mathcal{L}_3 = \int du~ u^5 T_{uu}^{(2)}$ as the first operator with a divergent one-point function. However, we expect that all of the ${\rm w}_{1+\infty}$ generators have finite one-point functions and are directly related to the universal soft factors in soft graviton theorems. }  Although only the primary descendants of the generators $\mathcal{W}^{\frac{3+\beta}{2}}$ have local expressions in terms of null integrals of the stress tensor when $\beta \geq 3$, interestingly, we find that their one-point functions also involve $\beta-2$ derivatives in $\bar{z}$.  Stripping off these derivatives on each side and transforming to momentum space (under which $x_1 \to i \partial_{p_1}$ and $x_3 \to - i \partial_{p_3}$), we find for all $\beta \geq 0$
\begin{equation} \label{eq:1ptresult}
    \begin{split}
        \langle \mathcal{O}(p_1) | \mathcal{W}^{\frac{3+\beta}{2}}(z,\bar{z}) |\mathcal{O}(p_3)\rangle
        &= \frac{1}{4 \pi } {\beta+2 \choose 2}    \frac{(-p_1^2)^2 \left(  \partial_z \hat n^\mu \hat  n^\nu \mathcal{L}_{1\mu\nu}\right)^\beta}{(-\hat n \cdot p_1)^{\beta+3}}   \langle \mathcal{O}(p_1) |\mathcal{O}(p_3)\rangle,
    \end{split}
\end{equation}
where here we have written our final result compactly in terms of the orbital angular momentum operator 
\begin{equation}
    \mathcal{L}_{k\mu\nu} = -i \left( p_{k\mu} \frac{\partial}{ \partial p^\nu_k}-p_{k\nu} \frac{\partial}{ \partial p^\mu_k} \right). 
\end{equation}

To illustrate the connection of \eqref{eq:1ptresult} to universal soft factors in soft graviton theorems, we first recall the statement of these theorems.  Given an expansion of a tree-level scattering amplitude in the energy of an emitted graviton, 
\begin{equation}
    \begin{split}
        \mathcal{A}_{n+1} (\omega \hat n; p_1, \cdots ,p_n)
             = \frac{\kappa}{2} \sum_{\beta = 0}^\infty \omega^{\beta-1} \mathcal{A}_{n+1}^{(\beta)} ( \hat n; p_1, \cdots ,p_n),
    \end{split}
\end{equation}
the soft graviton theorems \cite{Weinberg:1965nx,Cachazo:2014fwa,Li:2018gnc,Hamada:2018vrw,Guevara:2019ypd,Bautista:2019tdr,Himwich:2023njb} state that each term in the expansion takes the form 
\begin{equation}\label{subnleading}
    \begin{split}
        \mathcal{A}_{n+1}^{(\beta)}(\hat n; p_1,\cdots,p_n) & = \sum_{k = 1}^n S^{(\beta)}_k(\hat n)
        \mathcal{A}_n( p_1, \cdots ,p_n) +\varepsilon^-_{\mu\nu} \mathcal{B}^{\mu\nu}_{(\beta)} (\hat n; p_1, \cdots, p_n), 
    \end{split}
\end{equation}
where $\varepsilon^-_{\mu\nu}$ is the polarization of the graviton (here negative helicity), $\mathcal{A}_n$ is an $n$-point scattering amplitude without the emitted graviton, $\mathcal{B}^{\mu\nu}_{(\beta)}$ are ``non-universal" contributions that are known to arise generically for $\beta \geq 2$, and $S_k^{(\beta)}$ is the soft factor associated to the sub${}^\beta$-leading soft graviton theorem. $S_k^{(\beta)}$ transforms like an ${\rm SL}(2, \mathbb{C})$ primary of weight $(h, \bar{h}) = \left(\frac{-1-\beta}{2}, \frac{3-\beta}{2}\right)$.   For the purpose of comparison, we note that $\partial_{z}^{\beta+2} S^{(\beta)}_k$ is the primary descendant of the soft factor and takes the form \cite{Himwich:2023njb}\footnote{ This is the expression for  a soft factor for \textit{massive} momenta $p_k$, which was first considered in relation to asymptotic symmetries associated to leading and subleading soft graviton theorems by \cite{Campiglia:2015kxa}. Note that $(-\hat{n}\cdot p_k)^{-\Delta_{\rm CFT_2}}$ is a bulk-to-boundary propagator on the Euclidean AdS$_3$ resolution of timelike infinity. In \eqref{eq:twopt} the limit $\Delta \to 1$ sets $p_k$ to be null, and in this limit \eqref{soft-factor-exp} localizes to derivatives of delta functions, also presented in \cite{Himwich:2023njb}. } 
\begin{equation} \label{soft-factor-exp}
    \begin{split}
        \partial_{z}^{\beta+2} S^{(\beta)}_k (\hat{n})
            &= i^{\beta+2} {\beta+2 \choose 2} \frac{p_k^4 \left(\partial_z \hat n^\mu \hat n^\nu \mathcal{L}_k{}_{\mu\nu} \right)^{\beta}}{(-\hat n \cdot  p_k)^{\beta+3}}. 
    \end{split}
\end{equation}
Up to factors of $i$ and $4 \pi$, we note that this \emph{precisely} matches the form of the one-point functions.  Indeed, these missing factors should arise naturally from an asymptotic symmetry perspective upon turning the soft theorems into conservation statements for properly normalized Hermitian charges.

\section{Discussion} \label{sec:discussion}

\subsection{Summary of Forthcoming Extended Paper}

The present paper will be followed by the forthcoming longer work \cite{upcoming}, which will include a full discussion of the results presented here and a few extensions. In particular, it will include a full derivation of the basis of light-ray operators at fixed scaling dimension $\Delta$, their conformal transformations, a full enumeration of the ${\rm SL}(2,\mathbb{C})$ primaries at each $\Delta$ and their 4D descendant structure,  and the proof of \eqref{eq:localW} and \eqref{eq:localS}, including all of the necessary assumptions. The proof will include a derivation of the general $\rm{SL}(2,\mathbb{C})$ constraints on commutators and an expanded treatment of the structure of terms outside the  wedge.  The results and proof will also be generalized to include the adjoint action of $\mathcal{W}^p$ on $\mathcal{S}^{q,a}$, which results in the algebra \eqref{eq:adjointws}. It will also include details of the calculations of the one-point functions in Section \ref{sec:1pt} as well as their gauge-theoretic analogs. Finally, it will provide explicit calculations of the algebra \eqref{eq:localW} and \eqref{eq:localS} for the case of conformally coupled free scalars.

The expressions in the present paper all assume conformal symmetry, but they can, to varying extents, be generalized to non-conformal theories. \cite{upcoming} will present expressions in non-conformal theories for the covariant primaries $\mathcal{W}^{p}$ built from components of the stress tensor where tracelessness is not assumed. We will also discuss the extent to which the corresponding light-ray operator algebras can be constrained in generic quantum field theories.  A related question is what happens when moving the integral over null infinity into the bulk. While null infinity is certainly special in quantum field theories and different null surfaces have different physics (see \cite{Li:2025knf} for a recent discussion in the context of light-ray operators in the $O(N)$ model), in conformal theories different null surfaces can be related by conformal transformations \cite{Hofman:2008ar}. \cite{upcoming} will discuss the transformation of our operators on different null surfaces in conformal theories. Our explanation of the proof of \eqref{eq:localW} and \eqref{eq:localS} will also comment on the relation of $\mathcal{W}^{\frac{3}{2}}$, $\mathcal{W}^{2}$, and $\mathcal{S}^{1,a}$ to the generators of global symmetries. 

Finally, \cite{upcoming} will include a limited number of full commutators with terms outside the wedge and commutators of operators than $\mathcal{W}^p$ and $\mathcal{S}^{p,a}$. It may also include limited examples of direct calculations of the algebra from OPEs. 

\subsection{Future Directions}

Beyond broadly connecting the two vibrant research programs in celestial holography and light-ray operators, our work opens a number of exciting directions for future investigation.  First, it would be interesting to determine the general implications of this universal new symmetry algebra in the context of conformal field theory.  For example, there should be a generalization of the symmetry algebra to Lorentzian conformal field theories in any number of spacetime dimensions $D > 2$.  In addition, the symmetry algebra should imply novel constraints on correlation functions, for example in the form of sum rules (following  \cite{Kologlu:2019bco,Caron-Huot:2020adz,Chang:2023szz,Caron-Huot:2021enk}) or positivity constraints. Finally, one could work out explicit forms of the symmetry generators in known interacting conformal field theories such as $\mathcal{N}=4$ super-Yang-Mills and determine specific implications of the symmetry algebra for these theories. 

Second, it would be interesting to study this new universal symmetry algebra from the perspective of the AdS/CFT correspondence.  In particular, one could determine the geometric realization of the symmetry algebra in ${\rm AdS}_5$ (following \cite{Hofman:2008ar,Belin:2020lsr})  and perhaps exploit it to form novel generalizations of the ``stringy equivalence principle'' proposed in \cite{Kologlu:2019bco}.   

Third, it seems promising to explore the prospects for extending this symmetry algebra to relativistic \emph{quantum} field theories, i.e.~in the absence of full conformal symmetry.  Specifically, it would be interesting to study how the light-ray operators flow under renormalization and perhaps leverage these results to uncover more granular details about the flow of the $a$ and $c$ anomalies under renormalization \cite{CARDY1988749,OSBORN198997,Adams:2006sv,Komargodski:2011vj,Casini:2017vbe,Hartman:2023qdn,Hartman:2023ccw,Hartman:2024xkw}. One can also explore the possibility of new QCD observables or additional constraints on existing QCD observables that are implied by these universal symmetry algebras, especially those related to soft theorems at subleading orders (which for instance play an important role in \cite{Chang:2025zib}). 

Finally, this algebra and its construction may be used to uncover new lessons for celestial holography.   Specifically, our analysis in Section \ref{sec:1pt}, together with the results in the existing literature \cite{Cordova:2018ygx,Hu:2022txx,Hu:2023geb,Freidel:2021ytz,Freidel:2023gue}, strongly suggest that our universal class of light-ray operators can be identified with the matter contributions to the ``hard charges'' that generate asymptotic symmetries.  Since there are no closed-form formulas for these contributions to the charges in gravity, it would interesting to make this relation precise through a study of an asymptotic expansion of the Einstein equation.  There is also an intriguing \emph{mismatch} between the one-point functions in spinning operator states and soft factors for higher spinning particles.  It would interesting to find a way to exploit this mismatch to establish new constraints either in conformal field theory or quantum gravity in asymptotically flat spacetimes. Finally, the flow of the light-ray generators of $S$ and ${\rm w}_{1+\infty}$ under renormalization may shed new insight into the persistence of and corrections to the asymptotic symmetry algebras in quantum theories of gauge bosons and gravitons.

\section*{Acknowledgments} 

We are grateful to Jan Albert, Gr\'egoire Mathys, Noah Miller, Prahar Mitra, Ian Moult, and Ana-Maria Raclariu for useful conversations.   This work was completed with the support of NSF grant 2310633.  EH is supported by the John Archibald Wheeler Fellowship at the Princeton Center for Theoretical Science.

\begin{appendix}

\section{Large-$r$ Expansion and Conservation Equations}

In the flat Bondi coordinates \eqref{line-element-urz}, there are  nontrivial Christoffel coefficients
\begin{equation}
  \Gamma^{u}_{z\bar{z}} = r, \ \ \ \Gamma^{z}_{zr} = \frac{1}{r}, \ \ \ \Gamma^{\bar{z}}_{\bar{z}r} = \frac{1}{r}. 
\end{equation}
Conservation of the stress tensor in these coordinates becomes
\begin{equation} \label{eq:currentcons}
  \begin{aligned}
    0 &= g^{\mu \alpha}\nabla_{\alpha}T_{\mu \nu} = - 2 \nabla_r T_{u\nu} - 2 \partial_u T_{r\nu}  + \frac{2}{r^2}\nabla_z T_{\bar{z}\nu} + \frac{2}{r^2}\nabla_{\bar{z}} T_{z\nu}, 
  \end{aligned}
\end{equation}
which implies
\begin{align}
    0 &= (n-2)T_{uu}^{(n)} - \partial_u T_{ru}^{(n+1)} + \partial_zT_{\bar{z}u}^{(n-1)} + \partial_{\bar{z}}T_{zu}^{(n-1)}, \label{eq:current-conservation-u} \\ 
    0 &= (n-2)T_{uz}^{(n)} - \partial_u T_{rz}^{(n+1)} + \partial_zT_{z\bar{z}}^{(n-1)} + \partial_{\bar{z}}T_{zz}^{(n-1)}, \label{eq:current-conservation-z} \\ 
    0 &= (n-2)T_{ur}^{(n)} - \partial_u T_{rr}^{(n+1)} + \partial_zT_{\bar{z}r}^{(n-1)} + \partial_{\bar{z}}T_{zr}^{(n-1)} - 2 T_{z\bar{z}}^{(n-2)}. \label{eq:current-conservation-r}
\end{align}
In general quantum field theories we assume the falloffs 
\begin{equation}
    \begin{split} 
        T_{uu},T_{uz}, T_{zz} \sim \frac{1}{r^2},\quad \quad T_{z \bar{z}} \sim \frac{1}{r}, \quad \quad  T_{rz} \sim \frac{1}{r^3}, \quad \quad T_{ru}, T_{rr} \sim \frac{1}{r^4}. 
    \end{split}
\end{equation}

\subsection{Simplifications in Conformal Theories} \label{sec:conformalsimp}

In conformal field theories, tracelessness of the stress tensor implies
\begin{equation}
  T_{\mu}^{\mu} = - 4 T_{ur} + \frac{4}{r^2}T_{z\bar{z}},
\end{equation}
which gives 
\begin{equation}
T_{ur}^{(n)} = T_{z\bar{z}}^{(n-2)}. 
\end{equation}
This implies stricter falloff conditions on the stress tensor as follows. Tracelessness directly implies $T_{z\bar{z}}^{(1)} = 0$ because $T_{ur}^{(3)} = 0$.  Also, then $T_{rz}^{(3)} = 0$ by \eqref{eq:current-conservation-z} with $n=2$ and $T_{rr}^{(4)} = 0$ by \eqref{eq:current-conservation-r} with $n=3$.  With these simplifications, \eqref{eq:current-conservation-r} with $n=4$ simplifies to 
\begin{equation}
    \begin{split}
         0&= -\partial_u T_{rr}^{(5)} + 2T_{ur}^{(4)} - 2T_{z \bar{z}}^{(2)}
         =-\partial_u T_{rr}^{(5)},
    \end{split}
\end{equation}
where in the last equality we enforced the tracelessness condition.  Therefore $T_{rr}^{(5)}$ is also zero in a conformal field theory, and we find the falloffs \eqref{cft-fall-offs}.

\subsection{Spin-one Conserved Currents}

Conservation of global symmetry currents (with gauge indices suppressed) implies
\begin{equation} \label{eq:jcons}
  \begin{aligned}
    0 = g^{\mu\alpha}\nabla_{\alpha} j_{\mu} =  (n-2)j_u^{(n)} - \partial_u j_r^{(n+1)} + \partial_z j_{\bar{z}}^{(n-1)} + \partial_{\bar{z}}j_z^{(n-1)}.
  \end{aligned}
\end{equation}
In general quantum field theories we assume the falloffs 
\begin{equation} \label{eq:currentfalloffs}
j_u, j_z, j_{\bar{z}} \sim \frac{1}{r^2},\quad \quad \quad  j_{r} \sim \frac{1}{r^4}. 
\end{equation}

\section{4D Conformal Algebra}

The global generators of conformal symmetry can be expressed covariantly in terms of the stress-energy tensor as 
\begin{equation}
    \begin{split}
        Q_\zeta = \int_{\Sigma} d \Sigma^\rho ~\zeta^\mu T_{\rho \mu}, 
    \end{split}
\end{equation}
where $\zeta$ is an appropriately chosen vector field. In Cartesian coordinates, the vector fields that generate the 4D conformal transformations are given by
\begin{equation}
    \begin{split}
        P_\mu:&\quad \zeta_{(\mu)} = \partial_\mu, \\
        J_{\mu\nu}:&\quad \zeta_{(\mu \nu)} = x_\mu \partial_\nu -x_\nu \partial_\mu, \\
        D:& \quad \zeta = x^\mu \partial_\mu, \\
        K_\mu: & \quad \zeta_{(\mu)} =  2 x_\mu x^\nu \partial_\nu - x^2 \partial_\mu.  
    \end{split}
\end{equation}
The algebra of the global charges is 
\begin{equation}
\begin{aligned}
\left[ J_{\mu\nu}, J_{\alpha \beta}\right] &= -i \left(\eta_{\mu \beta}J_{\nu \alpha} + \eta_{\nu\alpha}J_{\mu\beta} - \eta_{\mu\alpha}J_{\nu\beta}-\eta_{\nu\beta}J_{\mu\alpha}\right), \\
\left[J_{\mu\nu},P_{\alpha}\right] &= -i\left(\eta_{\nu\alpha}P_{\mu} - \eta_{\mu\alpha}P_{\nu}\right),\ \ \qquad \left[J_{\mu\nu},K_{\alpha}\right] = -i\left(\eta_{\nu\alpha}K_{\mu} - \eta_{\mu\alpha}K_{\nu}\right), \\
\left[D,P_{\mu}\right] &= - i P_{\mu}, \ \quad \left[D,K_{\mu}\right] = i K_{\mu}, \quad \left[P_{\mu}, K_{\nu}\right] = -i \left(2 \eta_{\mu\nu}D + 2 J_{\mu\nu}\right),\\
\left[J_{\mu\nu},D\right] &= \left[D,D\right] = \left[P_{\mu},P_{\nu}\right] = \left[K_{\mu},K_{\nu}\right] = 0.
\end{aligned} 
\end{equation}
The action of the global charges on operators is
\begin{align}
        \left[P_\mu, \mathcal{O}(x)\right] &= - i \partial_{\mu} \mathcal{O}(x) ,\label{eq:Paction} \\
        \left[J_{\mu\nu}, \mathcal{O}(x)\right] &= - 2 i x_{\small[\mu} \partial_{\nu\small]} \mathcal{O}(x) + M_{\mu\nu}\cdot O(x),\\
        \left[D, \mathcal{O}(x)\right] &= -i x^\mu \partial_\mu \mathcal{O}(x) - i \Delta \mathcal{O}(x), \\
        \left[K_\mu, \mathcal{O}(x)\right] & =  -i2 x_\mu x^\nu \partial_\nu\mathcal{O}(x) +i x^2 \partial_\mu \mathcal{O}(x) - 2i \Delta x_{\mu}\mathcal{O}(x) + 2 x^{\nu} M_{\mu\nu} \cdot \mathcal{O}(x),  
\end{align}
where Lorentz spin indices are suppressed and $M_{\mu\nu}$ for a fixed $\mu\nu$ is a matrix in the Lorentz representation of $\mathcal{O}(x)$. Their action on light-ray operators will be provided in \cite{upcoming}. 

In flat Bondi coordinates, the vector fields that generate conformal transformations become the following. For translations, 
\begin{equation}
\zeta_P[f] = f \partial_u - \frac{1}{r}\left(\partial_z f \partial_{\bar{z}} + \partial_{\bar{z}}f \partial_z\right) + \partial_z \partial_{\bar{z}} f \partial_r 
\end{equation}
for $f = 1, z, \bar{z}, z \bar{z}$.  
For Lorentz transformations, 
\begin{equation}
  \zeta [Y] = \frac{u}{2}\left(\partial_z Y^z + \partial_{\bar{z}} Y^{\bar{z}} \right)\partial_u -\frac{r}{2}\left(\partial_z Y^z + \partial_{\bar{z}} Y^{\bar{z}} \right)\partial_r +\left(Y^z - \frac{u}{2r} \partial_{\bar{z}}^2 Y^{\bar{z}}\right)\partial_z + \left(Y^{\bar{z}} - \frac{u}{2r} \partial_{z}^2 Y^{z}\right)\partial_{\bar{z}}.
\end{equation}
For dilations,
\begin{equation}
  x^{\mu}\partial_{\mu} = u\partial_u + r\partial_r,
\end{equation}
and for special conformal transformations,  using $x^2 = -ur$ gives 
\begin{equation}
\zeta_K[f] = -\left(\partial_z \partial_{\bar{z}}f\right)u^2 \partial_u - u \left(\partial_z f \partial_{\bar{z}} + \partial_{\bar{z}} f \partial_z \right) - f r^2 \partial_r 
\end{equation}
for $f = 1, z, \bar{z}, z \bar{z}$. 
Thus we find the charges on null infinity
\begin{align}
    Q_{P[f]} &= \int du d^2z~ f T_{uu}^{(2)} \label{eq:globalP}, \\
    Q_{J[Y]} &= \int du d^2z \left(\frac{u}{2}(\partial_AY^A)T_{uu}^{(2)} + Y^AT_{uA}^{(2)}\right) \label{eq:globalJ},\\
    Q_{D} &= \int du d^2z~ u T_{uu}^{(2)} \label{eq:globalD} ,\\
    Q_{K[f]} &= \int du d^2z ~ f \left(u^2 \partial_z\partial_{\bar{z}}T_{uu}^{(2)} + T_{z\bar{z}}^{(2)} - u \partial_z T_{u\bar{z}}^{(2)} - u \partial_{\bar{z}}T_{uz}^{(2)} \right) ,\label{eq:globalK}
\end{align}
for $f = 1, z, \bar{z}, z\bar{z}$, and $Y^z = 1, z, z^2$, and we have used tracelessness  $T_{ur}^{(4)} = T_{z\bar{z}}^{(2)}$ to simplify $Q_{K[f]}$.

\end{appendix}

\bibliography{w-infinity-in-CFT}
\bibliographystyle{utphys}

\end{document}